\documentclass[conference,twocolumn]{IEEEtran}

\usepackage{subfigure}
\usepackage{graphicx}
\usepackage{dcolumn}
\usepackage{bm}
\usepackage{amsfonts}
\usepackage{amsmath}
\usepackage{amssymb}
\usepackage{amsthm}
\usepackage{citesort}

\ifCLASSINFOpdf

\else

\fi

\begin{document}
%
\title{Strong Spatial Mixing and Approximating Partition Functions of Two-State Spin Systems without Hard Constrains}

\author{\IEEEauthorblockN{\large \textsf{Jinshan Zhang}}
\IEEEauthorblockA{\textsf{Department of} \\\textsf{Mathematical Sciences}\\
\textsf{Tsinghua University}\\
\textsf{Beijing, China  100084} \\
\textsf{Email: zjs02@mails.tsinghua.edu.cn}}}

\maketitle

\textbf{Abstract:} We prove Gibbs distribution of two-state spin
systems(also known as binary Markov random fields) without hard
constrains on a tree exhibits strong spatial mixing(also known as
strong correlation decay), under the assumption that, for arbitrary
`external field', the absolute value of `inverse temperature' is
small, or the `external field' is uniformly large or small. The
first condition on `inverse temperature' is tight if the
distribution is restricted to ferromagnetic or antiferromagnetic
Ising models.

Thanks to Weitz's self-avoiding tree, we extends the result for
sparse on average graphs, which generalizes part of the recent work
of  Mossel and Sly\cite{MS08}, who proved the strong spatial mixing
property for ferromagnetic Ising model. Our proof yields a different
approach, carefully exploiting the monotonicity of local recursion.
To our best knowledge, the second condition of `external field' for
strong spatial mixing in this paper  is first considered and stated
in term of `maximum average degree' and `interaction energy'. As an
application, we present an FPTAS for partition functions of
two-state spin models without hard constrains under the above
assumptions in a general family of graphs including interesting
bounded degree graphs.

 \IEEEpeerreviewmaketitle

\section{Introduction}
Counting problem has played an important role in theoretic computer
science since Valiant\cite{Va79} introduced $\#$P-Complete
conception and proved many enumeration problems are computationally
intractable. The most successful and powerful existing method for
counting problem is due to Markov Chain method, which has been
successfully used to provide a fully polynomial randomized
approximation schemes (FPRAS)(which approximates the real value
within a factor of $\epsilon$ in polynomial time of the input and
$\epsilon^{-1}$ with the probability $\geq$ 3/4) for convex
bodies\cite{DFK91} and the number of perfect matchings on bipartite
graphs\cite{JSV04}. Since many counting problems such as the number
of matchings, independent sets, circuits\cite{MS06} etc. can be
viewed as special cases of computing partition functions associated
with Gibbs measures in statistical physics. Hence studying the
computation of partition function is a natural extension of counting
problems.

Self-reducing \cite{JVV86} or conditional probability method is a
well known method to compute partition functions if the marginal
probability of a vertex can be efficiently approximated. Gibbs
sampling also known as Glauber dynamics is a popular used method to
approximate marginal probability. This is a Markov Chain approach
locally updating the chain according to conditional Gibbs measure.
Hence studying the convergence rate(also known as mixing time) of
Glauber dynamics becomes a major research direction. Recently the
problem whether the Glauber dynamics converges `fast'(in a
polynomial time of the input and logarithm of reciprocal of sampling
error) deeply related to whether a phase transition takes place in
statistical model has been extensively studied, see \cite{MWW07} for
hard core model(also known as independent set model) and
\cite{MS08}\cite{GM07} for ferromagnetic Ising model. Another
approach to approximate marginal probability comes from the property
of the structure of Gibbs measures on various graphs. This method
utilizes local recursion and leads to deterministic approximation
schemes rather than random approximation schemes of Markov Chain
method. Our paper focuses on this recursive approach.

The recursive approach for counting problems is introduced by
Weitz\cite{We06} and Bandyopadhyay, Gamarnik \cite{BG06} for
counting the number of independent sets and colorings. The key of
this method is to establish the $strong$ $spatial$ $mixing$ property
also known as $strong$ $correlation$ $decay$ on certain defined
rooted  trees, which means the marginal probability of the root is
asymptotically independent of the configuration on the leaves far
below. Usually the exponential decay with the distance implies a
deterministic polynomial time approximating algorithm for marginal
probability of the root. In \cite{We06}, Weitz establishes the
equivalence between the marginal probability of a vertex in a
general graph $G$ and that of the root of a tree named
$self$-$avoiding$ $tree$ associated with $G$ for two-state spin
systems and shows the correlations  on any graph decay at least as
fast as its corresponding self-avoiding tree. He also proves the
strong correlation decay for hard-core model on bounded degree
trees. Later Gamarnik et.al.\cite{GK07} and Bayati
et.al.\cite{BGKNT06} bypass the construction of a self-avoiding
tree, by instead creating a certain $computation$ $tree$ and
establishing the strong correlation decay on the corresponding
computation tree for list coloring and matchings problems.  An
interesting relation between self-avoiding tree and computation tree
is that they share the same recursive formula for hard-core model.
Considering the motivation of construction of the self-avoiding
tree, Jung and Shah\cite{JS06} and Nair and Tetali \cite{NT07}
generalize Weitz's work for certain Markov random field models, and
Lu et.al.\cite{LMM07} for TP decoding problem. Mossel and
Sly\cite{MS08} show ferromagnetic Ising model exhibits strong
correlation decay on `sparse on average' graph under the tight
assumption that the `inverse temperature' in term of `maximum
average degree' is small.

In this paper, based on self-avoiding tree, we establish the strong
spatial mixing for general two-state spin systems also know binary
Markov random field without hard constrains on a graph that are
sparse on average under certain assumptions.  Our first condition is
on the `inverse temperature'. We show that there exits a value $J_d$
in term of `maximum average degree' $d$, if the absolute value of
the `inverse temperature' is smaller than $J_d$, for arbitrary
`external field', the Gibbs measure exhibits strong spatial mixing
on a sparse on average graph. Since for (anti)ferromagnetic Ising
model, strong spatial mixing on a finite regular tree implies
uniqueness of Gibbs measures of infinite regular
tree\cite{Ge88}\cite{We05}. $J_d$ in our setting is the critical
point for uniqueness of Gibbs measures of infinite regular tree with
degree of each vertex $d$, implying our condition is also necessary
on trees. The condition is the same as that of Mossel and
Sly\cite{MS08} when ferromagnetic Ising model is the only focus.
This makes part of their work in our framework. Our proof yields a
different approach, and also avoids the argument between weak
spatial mixing and strong spatial mixing employed in \cite{We06}. In
fact our proof is based an inequality similar to the one in
\cite{Ly89} and carefully exploits monotonicity of the recursive
formula. The recursive formula on trees is well known. Recently
Pemantle and Peres \cite{PP06} use it to present the exact capacity
criteria that govern behavior at critical point of ferromagnetic
Ising model on trees under various boundary conditions. Our second
condition is for `external field'. We prove for any `inverse
temperature' on a graph which are sparse on average Gibbs
distribution exhibits strong spatial mixing when the `external
field' is uniformly larger than $B(d,\alpha_{\max},\gamma)$ or
smaller than $-B(d,-\alpha_{\min},\gamma)$, where $d$ is `maximum
average degree' and $\alpha_{\min}$, $\alpha_{\max}$, $\gamma$ are
parameters of the system. To our best knowledge, this condition on
`external field' is first considered for strong spatial mixing. The
technique employed in the proof is Lipchitz approach which has been
used in \cite{BG06}\cite{BGKNT06}\cite{GK07}. The novelty here is
that we propose a `path' characterization of this method, allowing
us to give the `external field' condition in term of `maximum
average degree' rather than maximum degree. Some notations of the
`sparse on average' graphs have appeared in \cite{MS08}. These are
graphs where the sum degrees along each self-avoiding path(a path
with distinct vertices) with length $O($log$n)$ is $O($log$n)$.

As an application of our results, we present a fully polynomial time
approximation schemes(FPTAS)(which approximates the real value
within a factor of $\epsilon$ in polynomial time of the input and
$\epsilon^{-1}$) for partition functions of two-state spin systems
without hard constrains under our assumptions on the graph
$G=(V,E)$, where, for each vertex $v\in V$ of $G$, the number of
total vertices of its associated self-avoiding tree $T_{saw(v)}$
with hight $O($log$n)$ is $O(n^{O(1)})$. This includes bounded
degree graph and especially $Z^{d}$ lattice more concerned in
statistical physics. Jerrum and Sinclair \cite{JS93} provided an
FPRAS for partition function of ferromangetic Ising model for any
graph with any positive `inverse temperature' and identical external
field for all the vertices. Their results do not include the case
where different vertices have different external field, and are not
applied to antiferromagnetic Ising model where the `inverse
temperature' is negative either.

The remainder of the paper has the following structure. In Section
II, we present some preliminary definitions and main results. We go
on to prove the theorems in Section III. Section IV is devoted to
propose an FPTAS for the partition functions under our conditions.
Further work and conclusion are given in Section IV.

\section{Preliminaries and Main Results}
\subsection{Two-State Spin Systems}
In the two-state spin systems on a finite graph $G=(V,E)$ with
vertices $V=\{1,2,\cdots,n\}$ and edge set $E$, a configuration
consists of an assignment $\sigma=(\sigma_{i})$ of $\Omega=\{\pm1\}$
values, or ``spins", to each vertex(or``sites") of $V$. Each vertex
$i\in V$ is associated with a random variable $X_i$ with range
${\pm1}$. We often refer to the spin values $\pm1$ as $(+)$ and
$(-)$. The probability of finding the system in configuration
$\sigma\in\Omega^n$ is given by the joint distribution of $n$
 dimensional random vector $X=\{X_1,X_2,\cdots,X_{n}\}$(also known as the Gibbs
distribution with the nearest neighbor interaction)\\
\begin{displaymath} P_G(X=\sigma)=\frac{1}{Z(G)}\exp(\sum\limits_{(i,j)\in
E}\beta_{ij}(\sigma_i,\sigma_j)+\sum\limits_{i\in V}h_i(\sigma_i)).
\end{displaymath}
Here $Z(G)$ is called partition function of the system and a
normalized factor such that
$\sum_{\sigma\in\Omega^n}P_G(X=\sigma)=1$, and $h_i$ and
$\beta_{ij}$ are defined as a function from $\Omega$ and $\Omega^2$
to $R\cup\{\pm \infty\}$ respectively. We use notation
$\beta_{ij}(a,b)=\beta_{ji}(b,a)$. We say the system has hard
constraints if there exit an edge $(i,j)\in E$ or a vertex $k$, and
an assignment $\sigma_i$, $\sigma_j$ or $\sigma_k$ such that
$\beta_{ij}(\sigma_i;\sigma_j)=\infty$ or $h_k(\sigma_k)=\infty$
(e.g. hard-core model is one of the systems with hard constrains
where $\beta_{ij}(+,+)=-\infty$,
$\beta_{ij}(+,-)=\beta_{ij}(-,+)=\beta_{ij}(-,-)=0$ and $h_i(-)=0$).
In this paper we focus to the systems without hard constrains. We
call the function $\beta_{ij}$  `interaction energy' and $h_i$
`applied field' . If
$\beta_{ij}(\sigma_i,\sigma_j)=J_{ij}\sigma_i\sigma_j$ and
$h_i=B_i\sigma_i$ for all the edge $(i,j)\in E$ and vertex $i\in V$,
where $J_{ij}$ and $B_i$ are constant numbers varying with edges or
vertices, the system is called Ising model. Further, if $J_{ij}$ is
uniformly (negative)positive for all $(i,j)\in E$, the system is
called (anti)ferromagnetic Ising model. $J_{ij}$ and $B_i$ are
called $inverse$ $temperature$ and $external$ $field$ respectively.
To match the notation of Ising model, set
$J_{ij}=\frac{\beta_{ij}(+,+)+\beta_{ij}(-,-)-\beta_{ij}(-,+)-\beta_{ij}(+,-)}{4}$
and $B_i=\frac{h_i(+)-h_i(-)}{2}$ for all edges and vertices , in
this paper we call $J_{ij}$ and $B_i$ are `inverse temperature' and
`external field' of general two-state spin systems without hard
constrains (denoted by TSSHC for abbreviation)respectively. For any
$\Lambda\subseteq V$, $\sigma_{\Lambda}$ denotes the set $\{
\sigma_i, i\in \Lambda\}$. With a little abuse of notations,
$\sigma_{\Lambda}$ also denotes the configuration that  $i$ is fixed
$\sigma_i$, $\forall i\in \Lambda$. Let $Z(G,\Phi)$ denote the
partition function under the condition $\Phi$, e.g. $Z(G,X_1=+)$
represent the partition function under the condition the vertex $1$
is fixed $+$.

\subsection{Definitions and Notations}
\textbf{Definition 2.1} (Self-Avoiding Tree) Consider a graph
$G=(V,E)$ and  a vertex $v\in V$ in $G$. Given any order of all the
vertices in $G$. There is  associated partial order on $E$ of the
order on $V$ defined as $(i,j)>(k,l)$ iff $(i,j)$, $(k,l)$ share a
common vertex and $i+j > k+l$. The self-avoiding tree
$T_{saw(v)}(G)$(for simplicity denoted by $T_{saw(v)}$)
corresponding to the vertex $v$ is the tree of self-avoiding walks
originating at $v$ except that the vertices closing a cycle are also
included in the tree and are fixed to be either $+$ or $-$.
Specifically, the vertex of the $T_{saw(v)}$ closing a cycle is
fixed $+$ if the edge ending the cycle is larger than the edge
starting the cycle and $-$ otherwise. 
Given any configuration $\sigma_{\Lambda}$ of $G$, $\Lambda\subset
V$, the self-avoiding tree is constructed the same as the above
procedure except that the vertex  which is a copy of the vertex $i$
in $\Lambda$ is fixed to the same spin $\sigma_i$ as $i$ and the
subtree below it is not constructed(See Figure 1). Hence, for any
configuration $\sigma_{\Lambda}$ of $G$, $\Lambda\subset V$, we also
use $\sigma_{\Lambda}$ to denote the configuration of $T_{saw(v)}$
obtained by imposing the condition corresponding to
$\sigma_{\Lambda}$ as above. For any $(i,j)\in E$ and $i\in V$ of
$G$, the `interaction energy' function  and `applied field' function
on all their copies  of the induced system on $T_{saw(v)}$ by $G$
are the same as $\beta_{ij}$ and $h_i$
respectively.\\

We now provide the remarkable property of the self-avoiding tree,
one of two main results of \cite{We06}, which is one of the
essential techniques of
our proofs.\\
\begin{figure}[ht]
\centering
\includegraphics[scale=.35]{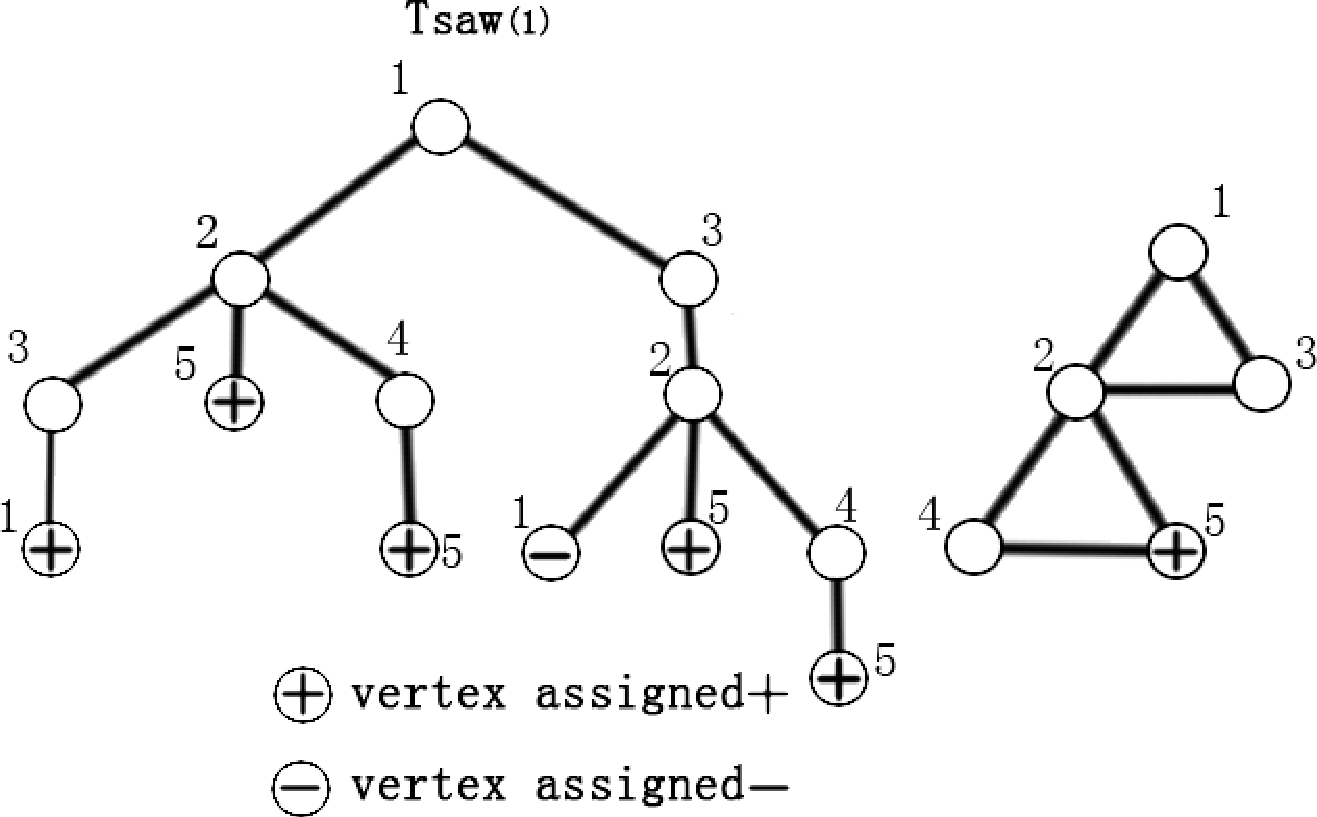}
\caption{\small{The graph with one vertex assigned + (Right)and its
corresponding self-avoiding tree $T_{saw(1)}$(Left)}}
\end{figure}
\textbf{Proposition 2.1}\emph{ For two-state spin systems on
$G=(V,E)$, for any configuration $\sigma_{\Lambda}$, $\Lambda\subset
V$ and any vertex $v\in V$, then
\begin{displaymath}
P_G(X_v=+|\sigma_{\Lambda})=P_{T_{saw(v)}}(X_v=+|\sigma_{\Lambda}).
\end{displaymath}}\\

In order to generalize our result to more general families of graphs
, which are sparse on average, we need some definitions and notation
of these graphs.\\

 \textbf{Definition 2.2} Let $|A|$ denote the cardinality of the set $A$.
 The length of a path is the number of
edges it contains.  The distance of two vertices in a graph is the
length of shortest path connecting these two vertices. A path
$v_1,v_2,\cdots$ is called a self-avoiding path if for all $i\neq j$
, $v_i\neq v_j$. In a graph $G=(V,E)$, let $d(u,v)$ denote the
distance between $u$ and $v$, $u$,$v\in V$. The distance between a
vertex $v\in V$ and a subset $\Lambda\subset V$ is defined by
$d(v,\Lambda)=$min$\{d(v,u):u\in\Lambda\}$. The set of vertices
within distance $l$ of $v$ is denoted by $V(G,v,l)=\{u: d(v,u)\leq
l\}$. Similarly, the set of vertices with distance  $l$ of $v$ is
denoted by $S(G,v,l)=\{u: d(v,u)= l\}$.
 We call a vertex  at the height $t$ of a rooted tree if the
distance between it and the root is $t$. Let $\delta_v$ denote the
degree of $v$ in $G$. The $maximal$ $path$ $density$ $m$ is defined
by $m(G,v,l)=\max\limits_{\Gamma}\sum\limits_{u\in \Gamma}\delta_u$,
where the maximum is taken over all self-avoiding paths $\Gamma$
starting at $v$ with length at most $l$. The $maximum$ $average$
$path$ $degree$ $\delta(G,v,l)$ is defined by
$\delta(G,v,l)=(m(G,v,l)-\delta_v)/l, l\geq 1$. The $maximum$
$average$ $degree$ of $G$ is defined by $\Delta(G,l)=\max_{v\in
V}\delta(G,v,l)$. Roughly speaking, in this paper, a family of
graphs $\mathcal{G}$ is sparse on average if there exits a constant
number $a$ and $d$ such
that $\Delta(G,a\log n)\leq d$ for any $G\in\mathcal{G}$.\\

Some properties of the above definitions are useful in our proof, we
present them. Most of proofs are simply obtained by induction and
can be found in \cite{MS08}.\\

\textbf{Proposition 2.2} \emph{ Let $j$, $l$ denote positive natural
numbers, then
\begin{displaymath}m(G,v,jl)\leq j\max\limits_{u\in
G}\{m(G,u,l)-\delta_u\}+\delta_v.\end{displaymath}}\\

\textbf{Proposition 2.3} \emph{Let $l$ be natural numbers, then
\begin{displaymath}
|S(T_{saw(v)},v,l+1)|\leq\delta_v(\delta(G,v,l)-1)^l.
\end{displaymath}}\\

\textbf{Proposition 2.4} \emph{Let $j$, $l$ be natural numbers, then
\begin{displaymath}
|V(T_{saw(v)},v,jl)|\leq (\max\limits_{u\in
V}|V(T_{saw(u)},u,l)|)^j.
\end{displaymath}}\\

\textbf{Definition 2.3} ((Exponential) Strong Spatial Mixing) Let
$G=(V,E)$ be a graph with $n$ vertices. The Gibbs distribution of
two-state spin systems on $G$ exhibits strong spatial mixing iff for
any vertex $v\in V$, subset $\Lambda\subset V$, any two
configurations $\sigma_{\Lambda}$ and $\eta_{\Lambda}$ on $\Lambda$,
denote $\Theta=\{v\in\Lambda: \sigma_v\neq \eta_v\}$ and
$t=d(v,\Theta)$,
\begin{displaymath}
|P_G(X_v=+|\sigma_{\Lambda})-P_G(X_v=+|\eta_{\Lambda})|\leq
f(t),\end{displaymath} where $f(t)$ goes to zero if $t$ goes to
infinity and is called decay function.\\
 For the purpose of our settings, we present a weak form
of exponential strong spatial mixing. We say the distribution
exhibits exponential strong spatial mixing if there exits positive
numbers $a$, $b$, $c$ independent of $n$ such that $f(t)= b\exp(-c
t)$ when $t=ka$log$n$, $k=1,2,\cdots$.

Remark: In the above definition of (exponential) strong spatial
mixing, $P_G(X_v=+|\sigma_{\Lambda})$ and
$P_G(X_v=+|\eta_{\Lambda})$ can be replaced by
$\log(P_G(X_v=+|\sigma_{\Lambda}))$ and
$\log(P_G(X_v=+|\eta_{\Lambda}))$ respectively if $d(v,\Lambda)$ is
large than a constant number,  due to the inequality
$2x\leq$log$(1+x)\leq x$ when $|x|\leq .5$, and we call it
 the logarithmic form exponential strong spatial mixing. )\\

\textbf{Definition 2.4} (FPTAS) An approximation algorithm is called
a fully polynomial time approximation scheme(FPTAS) iff for any
$\epsilon>0$, it takes a polynomial time of input and
$\epsilon^{-1}$ to output a value $\bar{M}$ satisfying
\begin{displaymath}
1-\epsilon\leq\frac{\bar{M}}{M}\leq1+\epsilon,
\end{displaymath}
 where $M$ is the real value.\\

Remark: In the above definition $1-\epsilon$ and $1+\epsilon$ can be
replaced by $e^{-\epsilon}$ and $e^{\epsilon}$.
\subsection{Main Results}
For simplicity , We use the following notations. Consider a
two-state spin systems with hard constrains(TSSHC) on a graph
$G=(V,E)$ with $n$ vertices $V=\{1,2,\cdots,n\}$ and edge set $E$.
Let $J=\max_{(i,j)\in E}|J_{ij}|$, $B_{\min}=\min_{i\in V}B_{i}$,
 $B_{\max}=\max_{i\in V}B_{i}$, $\alpha_{\max}=\max\limits_{(i,j)\in
E}\{\beta_{ij}(-,-)-\beta_{ij}(+,-),\beta_{ij}(-,+)-\beta_{ij}(+,+)\}$,
$\alpha_{\min}=\min\limits_{(i,j)\in
E}\{\beta_{ij}(-,-)-\beta_{ij}(+,-),\beta_{ij}(-,+)-\beta_{ij}(+,+)\}$,
$\gamma_{ij}=\max_{(i,j)\in
E}\{\frac{|b_{ij}c_{ij}-a_{ij}d_{ij}|}{a_{ij}c_{ij}},\frac{|b_{ij}c_{ij}-a_{ij}d_{ij}|}{b_{ij}d_{ij}}\}$,
$\gamma=\max_{(i,j)\in E}\{\gamma_{ij}\}$, where
$J_{ij}=\frac{\beta_{ij}(+,+)+\beta_{ij}(-,-)-\beta_{ij}(-,+)-\beta_{ij}(+,-)}{4}$,
$B_i=\frac{h_i(+)-h_i(-)}{2}$ are `inverse temperature' and
`external field' respectively, and $a_{ij}=\exp(\beta_{ij}(+,+))$,
$b_{ij}=\exp(\beta_{ij}(+,-))$, $c_{ij}=\exp(\beta_{ij}(-,+))$,
$d_{ij}=\exp(\beta_{ij}(-,-))$.\\

\textbf{Theorem 2.1} \emph{Let $G=(V,E)$ be a graph with vertices
$V=\{1,2,\cdots,n\}$, edges set $E$ and TSSHC on it.  If there exit
two positive numbers $a>0$ and $d>0$ such that $\Delta(G,a\log
n)\leq d$, and when
\begin{displaymath}
(d-1)\tanh{J}<1
\end{displaymath}
or equivalently $J<J_d=\frac{1}{2}\log(\frac{d}{d-2})$, then the
Gibbs distribution of TSSHC exhibits logarithmic form exponential
strong spatial mixing for arbitrary `external field', specifically,
for any $i\in V$, any two configurations $\sigma_{\Lambda}$ and
$\eta_{\Lambda}$ on $\Lambda$, denote $\Theta=\{j\in\Lambda:
\sigma_j\neq \eta_j\}$ and $t=d(i,\Theta)=ka\log n +1$,
$k=1,2,\cdots,$
\begin{displaymath}
|\log(P_G(X_i=+|\sigma_{\Lambda}))-\log(P_G(X_i=+|\eta_{\Lambda}))|\leq
f(t),
\end{displaymath}
where $f(t)=4J\delta_i((d-1)\tanh J)^{t-1}$.}\\

Remark: If the graph is bounded with the maximum degree $D$, then
$d$ can be replaced by $D$ while  for any $a>0$,  and $J_{ij}$ is
the `inverse temperature' in (anti)ferromagnetic Ising model, then
theorem 2.1 still holds and $J_D$ is the critical point for
uniqueness of Gibbs measures on a infinite tree with maximum degree
$D$\cite{Ly89}. Note the decay function is slight different from the
definition since $\delta_i$ may be $O(\log n)$, however, in this
case we can choose $k$ large enough independent of $n$ such that
$f(t)=e^{-bt}$ when $n$ is large, where $b$ is a positive number
independent of $n$, then replace $a$ by $ka$ as required. In fact
in the application of the algorithm, this is not important. \\

 \textbf{Theorem 2.2} \emph{ Let $G=(V,E)$ be a graph with
vertices $V=\{1,2,\cdots,n\}$, edges set $E$ and TSSHC on it.  If
there exit two positive numbers $a>0$ and $d>0$ such that
$\Delta(G,a\log n)\leq d$, and $(d-1)\tanh{J}\geq1$, and when
\begin{displaymath}
B_{\min}>B(d,\alpha_{\max},\gamma)\ \ \ \ \ \ or \ \ \ \ \
B_{\max}<-B(d,-\alpha_{\min},\gamma)
\end{displaymath}
where
$B(d,\alpha,\gamma)=\frac{(d-1)\alpha}{2}+\log(\frac{\sqrt{\gamma(d-1)}+\sqrt{\gamma(d-1)-4}}{2})$,
the Gibbs distribution of TSSHC exhibits exponential strong spatial
mixing, specifically, for any $i\in V$, any two configurations
$\sigma_{\Lambda}$ and $\eta_{\Lambda}$ on $\Lambda$, denote
$\Theta=\{j\in\Lambda: \sigma_j\neq \eta_j\}$ and
$t=d(i,\Theta)=ka\log n +1$, $k=1,2,\cdots,$
\begin{displaymath}
|P_G(X_i=+|\sigma_{\Lambda})-P_G(X_i=+|\eta_{\Lambda})|\leq f(t),
\end{displaymath}
where
$f(t)=\frac{\delta_i\gamma}{4}(\frac{(d-1)\gamma\exp(2B_{\min}-(d-1)\alpha_{\max})}{(1+\exp(2B_{\min}-(d-1)\alpha_{\max}))^2})^{t-1}$
or
$f(t)=\frac{\delta_i\gamma}{4}(\frac{(d-1)\gamma\exp(2B_{\max}-(d-1)\alpha_{\min})}{(1+\exp(2B_{\max}-(d-1)\alpha_{\min}))^2})^{t-1}$
 respectively.}\\

Remark: It's easy to check $\gamma\geq 4\tanh J$, hence in theorem
2.2, if  $(d-1)\tanh{J}\geq1$, then $\gamma(d-1)-4\geq 0$. As a
corollary of Theorem 2.2, from its proof in section III, we know if
the graph is bounded degree with maximum degree is $d$, the
condition for `external field' can be relaxed to
$B_i>B(d,\alpha_{\max},\gamma)$ or $
B_i<-B(d,-\alpha_{\min},\gamma)$ for any $i\in V$, which does not
require
that `external field' is uniformly large or uniformly small.\\

\textbf{Theorem 2.3} \emph{Let $G=(V,E)$ be a graph with $n$
vertices $V=\{1,2,\cdots,n\}$, edges set $E$ and TSSHC on it.
If there exit two positive numbers $a>0$ and $d>0$ such that for any
$i\in V$
\begin{displaymath}
V(T_{saw(i)},i,a\log{n})\leq (d-1)^{a\log{n}},
\end{displaymath}
where $|V(T_{saw(i)},i,l)|= \{j\in T_{saw(i)}:d(i,j)\leq l\}$, then
when $J<J_d$ or $J\geq J_d$, $B_{\min}>B(d,\alpha_{\max},\gamma)$ or
$B_{\max}<-B(d,-\alpha_{\min},\gamma)$, there exits an
FPTAS for partition function of TSSHC on $G$.}\\

\section{Proofs}
We now proceed to prove the theorems. One of the technical lemmas
for the theorem 2.1 is an inequality similar to \cite{Ly89}. We present it now.\\

\textbf{Lemma 3.1}\emph{ Let $a$, $b$, $c$, $d$, $x$, $y$ be
positive numbers and $g(x)=\frac{ax+b}{cx+d}$ and
$t=|\frac{\sqrt{ad}-\sqrt{bc}}{\sqrt{ad}+\sqrt{bc}}|$, then
\begin{displaymath}
\max(\frac{g(x)}{g(y)},
\frac{g(y)}{g(x)})\leq(\max(\frac{x}{y},\frac{y}{x}))^t.
\end{displaymath}}\\

\textbf{Proof:} Case 1. $ad\geq bc$. Since $g(x)=
\frac{ax+b}{cx+d}=\frac{a}{c}-\frac{ad-bc}{c(cx+d)}$ is an
increasing function, w.l.o.g. suppose $x\geq y$ and let $x=zy$,
where $z\geq1$, then
\begin{displaymath}
\begin{split}
\log(\frac{g(x)}{g(y)})&=\int^{z}_{1}\frac{d(\log(\frac{g(\alpha
y)}{g(y)}))}{d\alpha}d\alpha\\& =\int^{z}_{1} (\frac{ay}{a\alpha
y+b}-\frac{cy}{c\alpha
y+d})d\alpha\\
&=\int^{z}_{1} \frac{(ad-bc)y}{(a\alpha y+b)(c\alpha y+d)}d\alpha\\
&=\int^{z}_{1} \frac{(ad-bc)y}{(\sqrt{ac}\alpha y-\sqrt{bd})^2+(\sqrt{bc}+\sqrt{ad})^2\alpha y}d\alpha\\
&\leq\int^{z}_{1} \frac{(ad-bc)y}{(\sqrt{bc}+\sqrt{ad})^2\alpha
y}d\alpha =\frac{\sqrt{ad}-\sqrt{bc}}{\sqrt{ad}+\sqrt{bc}}\log z.
\end{split}
\end{displaymath}
Hence,
\begin{displaymath}
\max(\frac{g(x)}{g(y)}, \frac{g(y)}{g(x)})=\frac{g(x)}{g(y)}\leq
(\frac{x}{y})^t= (\max(\frac{x}{y},\frac{y}{x}))^t,
\end{displaymath}
where $t=\frac{\sqrt{ad}-\sqrt{bc}}{\sqrt{ad}+\sqrt{bc}}$.\\
Case 2. $ad\leq bc$. Similar to the first case, $g(x)$ is a
decreasing function, let $h(x)=1/g(x)$, then $h(x)$ is an increasing
function, w.l.o.g. suppose $x\geq y$, repeat the process of Case 1
for $\frac{h(x)}{h(y)}$, then
\begin{displaymath}
\frac{h(x)}{h(y)}\leq
(\frac{x}{y})^{\frac{\sqrt{bc}-\sqrt{ad}}{\sqrt{ad}+\sqrt{bc}}}.
\end{displaymath}
Hence,
\begin{displaymath}
\max(\frac{g(x)}{g(y)},
\frac{g(y)}{g(x)})=\frac{g(y)}{g(x)}=\frac{h(x)}{h(y)}\leq
(\frac{x}{y})^t= (\max(\frac{x}{y},\frac{y}{x}))^t,
\end{displaymath}
where $t=\frac{\sqrt{bc}-\sqrt{ad}}{\sqrt{ad}+\sqrt{bc}}$. \ \ \
$\Box$\\

\textbf{Lemma 3.2}\emph { Let $T=(V,E)$ be a tree rooted at $0$ with
vertices $V=\{0,1,2,\cdots,n\}$, edge set $E$ and TSSHC on it.
Suppose some vertices are fixed.  Let $T_k$ and $T_l$ be two
subtrees of $T$ including vertex $k$ and $l$ respectively by
removing an edge $(k,l)$ where $d(k,0)<d(l,0)$. The fixed vertices
remain fixed on $T_k$ and $T_l$. Then the probability of $X_0=+$ on
$T$ equals the probability of $X_0=+$ on the subtree $T_k$ except
changing the `external field' $h_k$ to certain value $h^{'}_k$.}\\

\textbf{Proof:} Let $\Omega_{T_l}$ denote the configuration spaces,
$E_l$ and $V_l$ the edge set and vertices on $T_l$. Setting
\begin{displaymath}
\begin{split}
h^{'}_k(\sigma_k)&=h_k(\sigma_k)+\\
&\log(\sum\limits_{\sigma\in\Omega_{T_l}}e^{
\beta_{kl}(\sigma_k,\sigma_l)+ \sum\limits_{(i,j)\in
E_l}\beta_{ij}(\sigma_i,\sigma_j)+\sum\limits_{i\in
V_l}h_i(\sigma_i)})
\end{split}
\end{displaymath}
completes the proof.\ \ \  $\Box$\\

With Lemma 3.1 and Lemma 3.2, we now proceed to prove (exponential)
strong spatial mixing property on trees.\\

\textbf{Theorem 3.1} \emph{Let $T$ be a tree rooted at $0$ with
vertices $V=\{0,1,2,\cdots,n\}$, edge set $E$ and TSSHC on it.  Let
$\Lambda\subset V$ , $\zeta_{\Lambda}$ and $\eta_{\Lambda}$ be any
two configurations on $\Lambda$. Let $\Theta=\{i: \zeta_i\neq
\eta_i, i\in\Lambda\}$, $t=d(0,\Theta)$ and $s=|S(T,0,t)|=|\{i:
 d(0,i)=t, i\in T\}|$. Then
\begin{displaymath}
\max(\frac{P_T(X_0=+|\zeta_{\Lambda})}{P_T(X_0=+|\eta_{\Lambda})}
,\frac{P_T(X_0=+|\eta_{\Lambda})}{P_T(X_0=+|\zeta_{\Lambda})})\leq
\exp(4Js(\tanh{J})^{t-1})
\end{displaymath}}\\

\textbf{Proof:} For any $i\in V$, let $T_i$ denote the subtree with
$i$ as its root and $Z(i)$  be the TSSHC induced on $T_i$ by $T$.
Noting $T_0$ is $T$.  To prove the theorem, it's convenient to deal
with the ratio
$\frac{P_T(X_0=+|\zeta_{\Lambda})}{P_T(X_0=-|\zeta_{\Lambda})}$
rather than $P_T(X_0=+|\zeta_{\Lambda})$ itself. Denote
$R^{\zeta_{\Lambda}}_i\equiv
\frac{P_{T_i}(X_i=+|\zeta_{\Lambda_{i}})}{P_{T_i}(X_i=-|\zeta_{\Lambda_{i}})}$
where $\zeta_{\Lambda_{i}}$ is the condition  by imposing the
configuration $\zeta_{\Lambda}$ on $T_i$, and note a simple relation
if $x_1, x_2 \in (0,1)$, then $\frac{x_1}{x_2}\geq 1$ iff
$\frac{x_1}{1-x_1}\geq \frac{x_2}{1-x_2}$, further
$\max\{\frac{x_1}{x_2},\frac{x_2}{x_1}\}\leq
\max\{\frac{x_1/(1-x_1)}{x_2/(1-x_2)},\frac{x_2/(1-x_2)}{x_1/(1-x_1)}\}$.
Hence replace $x_1$ and $x_2$ by $P_T(X_0=+|\zeta_{\Lambda})$ and
$P_T(X_0=+|\eta_{\Lambda})$, we need only to show
\begin{equation}
\max(\frac{R^{\zeta_{\Lambda}}_0}{R^{\eta_{\Lambda}}_0}
,\frac{R^{\eta_{\Lambda}}_0}{R^{\zeta_{\Lambda}}_0})\leq
\exp(4Js(\tanh{J})^{t-1}).
\end{equation}
Theorem 3.1 follows by
$\max(\frac{P_T(X_0=+|\zeta_{\Lambda})}{P_T(X_0=+|\eta_{\Lambda})}
,\frac{P_T(X_0=+|\eta_{\Lambda})}{P_T(X_0=+|\zeta_{\Lambda})}) \leq
\max(\frac{R^{\zeta_{\Lambda}}_0}{R^{\eta_{\Lambda}}_0}
,\frac{R^{\eta_{\Lambda}}_0}{R^{\zeta_{\Lambda}}_0})$.\\
We go on to prove (1) by induction on $t$. Before we doing this,
some trivial cases need to be clarified. We are interested in the
case $t\geq 1$ and $0$ is unfixed. Let $\Gamma_{kl}$ denote the
unique self-avoiding path from $k$ to $l$ on $T$. If $i$ is a leave
on $T$ and $d(0,i)<t$, where $t=d(0,\Theta)$. Define $U=\{j\in V:
j\in \Gamma_{0i}, \exists k\in S(T,0,t), s.t. j\in\Gamma_{0k}\}$.
Note $U\neq \emptyset$ since $0\in U$. Let $j_i\in U$ such that
$d(i,j_i)=d(i,U)$. By lemma 3.2, we can remove the subtree bellow
$j_i$ and change external field $h_{j_i}$  at $j_i$ to $h^{'}_{j_i}$
without changing the probability of $X_0=+$. More importantly, this
process removes at least one leave at the hight $< t$, and does not
remove any vertex at the hight $\geq t$. Thus, w.l.o.g. suppose $T$
is a tree rooted at $0$ where any leave on it at the height $\geq
t$. Let $0_1,0_2,\cdots,0_q$ be the neighbors connected to $0$.  A
trivial calculation then gives that
\begin{displaymath}
\begin{split}
&R^{\zeta_{\Lambda}}_0=\frac{Z(T_0,X_0=+,\zeta_{\Lambda})}{Z(T_0,X_0=-,\zeta_{\Lambda})}\\
&=\frac{e^{h_0(+)}\sum\limits_{\sigma\in\Omega_0}e^{\sum\limits_{i=1}^q(\beta_{00_i}(+,\sigma_{0_i})+\sum\limits_{(k,l)\in
T_i}\beta_{kl}(\sigma_k,\sigma_l)+\sum\limits_{k\in
T_i}h_k(\sigma_k))}}{e^{h_0(-)}\sum\limits_{\sigma\in\Omega_0}e^{\sum\limits_{i=1}^q(\beta_{00_i}(-,\sigma_{0_i})+\sum\limits_{(k,l)\in
T_i}\beta_{kl}(\sigma_k,\sigma_l)+\sum\limits_{k\in T_i}h_k(\sigma_k))}}\\
&=e^{2B_0}\prod\limits^{q}_{i=1}\frac{\sum\limits_{\sigma\in\Omega_{T_i}}e^{\beta_{00_i}(+,\sigma_{0_i})+\sum\limits_{(k,l)\in
T_i}\beta_{kl}(\sigma_k,\sigma_l)+\sum\limits_{k\in
T_i}h_k(\sigma_k)}}{\sum\limits_{\sigma\in\Omega_{T_i}}e^{\beta_{00_i}(-,\sigma_{0_i})+\sum\limits_{(k,l)\in
T_i}\beta_{kl}(\sigma_k,\sigma_l)+\sum\limits_{k\in
T_i}h_k(\sigma_k)}}\\
\end{split}
\end{displaymath}
\begin{equation}
\begin{split}
&=e^{2B_0}
\prod\limits^{q}_{i=1}\frac{a_iZ(T_{0_i},X_i=+,\zeta_{\Lambda_i})+b_iZ(T_{0_i},X_i=-,\zeta_{\Lambda_i})}
{c_iZ(T_{0_i},X_i=+,\zeta_{\Lambda_i})+d_i Z(T_{0_i},X_i=-,\zeta_{\Lambda_i})}\\
&=e^{2B_0} \prod\limits^{q}_{i=1}\frac{a_i
R^{\zeta_{\Lambda}}_{0_i}+b_i}{c_i R^{\zeta_{\Lambda}}_{0_i}+d_i}.
\end{split}
\end{equation}
where $B_0=\frac{h_0(+)-h_0(-)}{2}$, $a_i=e^{\beta_{00_i}(+,+)}$,
$b_i=e^{\beta_{00_i}(+,-)}$, $c_i=e^{\beta_{00_i}(-,+)}$,
$d_i=e^{\beta_{00_i}(-,-)}$ . Now we check the base case $t=1$ where
$R^{\zeta_{\Lambda}}_{0_i}, R^{\eta_{\Lambda}}_{0_i} \in
[0,+\infty]$, by the monotonicity of $\frac{a_i
R^{\zeta_{\Lambda}}_{0_i}+b_i}{c_i R^{\zeta_{\Lambda}}_{0_i}+d_i}$
and $\frac{a_i R^{\eta_{\Lambda}}_{0_i}+b_i}{c_i
R^{\eta_{\Lambda}}_{0_i}+d_i}$,
\begin{displaymath}
\begin{split}
\max(\frac{R^{\zeta_{\Lambda}}_0}{R^{\eta_{\Lambda}}_0}
,\frac{R^{\eta_{\Lambda}}_0}{R^{\zeta_{\Lambda}}_0})&\leq
\prod\limits^{q}_{i=1}\max(\frac{a_id_i}{b_ic_i},\frac{b_ic_i}{a_id_i})\leq
e^{4qJ}
\end{split}
\end{displaymath}
Hence $t=1$, (1) holds. Assume by induction that (1) holds for
$t-1$, and we will show it holds for $t$. Let
$s_i=|S(T_{0_i},0_i,t-1)|$, $i=1,2,\cdots,q$, still using above
recursive procedure, then
\begin{displaymath}
\begin{split}
\max(\frac{R^{\zeta_{\Lambda}}_0}{R^{\eta_{\Lambda}}_0}
,\frac{R^{\eta_{\Lambda}}_0}{R^{\zeta_{\Lambda}}_0})&\leq\prod\limits^{q}_{i=1}\max(\frac{\frac{a_i
R^{\zeta_{\Lambda}}_{0_i}+b_i}{c_i
R^{\zeta_{\Lambda}}_{0_i}+d_i}}{\frac{a_i
R^{\eta_{\Lambda}}_{0_i}+b_i}{c_i
R^{\eta_{\Lambda}}_{0_i}+d_i}},\frac{\frac{a_i
R^{\eta_{\Lambda}}_{0_i}+b_i}{c_i
R^{\eta_{\Lambda}}_{0_i}+d_i}}{\frac{a_i
R^{\zeta_{\Lambda}}_{0_i}+b_i}{c_i
R^{\zeta_{\Lambda}}_{0_i}+d_i}})\\
&\leq\prod\limits^{q}_{i=1}\max(\frac{R^{\zeta_{\Lambda}}_{0_i}}
{R^{\eta_{\Lambda}}_{0_i}},\frac{R^{\eta_{\Lambda}}_{0_i}}
{R^{\zeta_{\Lambda}}_{0_i}})^{|\frac{\sqrt{a_id_i}-\sqrt{b_ic_i}}
{\sqrt{a_id_i}+\sqrt{b_ic_i}}|}\\
&\leq\prod\limits^{q}_{i=1}\max(\frac{R^{\zeta_{\Lambda}}_{0_i}}
{R^{\eta_{\Lambda}}_{0_i}},\frac{R^{\eta_{\Lambda}}_{0_i}}
{R^{\zeta_{\Lambda}}_{0_i}})^{\tanh J}
\end{split}
\end{displaymath}
where the second inequality comes from the Lemma 3.1.  According to
the hypothesis of induction $\max(\frac{R^{\zeta_{\Lambda}}_{0_i}}
{R^{\eta_{\Lambda}}_{0_i}},\frac{R^{\eta_{\Lambda}}_{0_i}}
{R^{\zeta_{\Lambda}}_{0_i}}) \leq \exp(4Js_i(\tanh{J})^{t-2})$, it's
sufficient to show
\begin{displaymath}
\begin{split}
\max(\frac{R^{\zeta_{\Lambda}}_0}{R^{\eta_{\Lambda}}_0}
,\frac{R^{\eta_{\Lambda}}_0}{R^{\zeta_{\Lambda}}_0})&\leq\prod\limits^{q}_{i=1}\exp(4Js_i(\tanh{J})^{t-1})\\
&=\exp(4Js(\tanh{J})^{t-1})
\end{split}
\end{displaymath}
where the last equation follows by $\sum_{i=1}^qs_i=s$. This
completes the proof of Theorem 3.1. \ \ \ $\Box$\\

With Theorem 3.1 and self-avoiding tree, it's enough to prove
Theorem 2.1.\\

\textbf{Proof of Theorem 2.1:} Due to Proposition 2.1, the only
thing left is to verify $|S(T_{saw(i)}, i, t)|=\delta_i(d-1)^{t-1}$
when $t=d(i, \Theta)=ka\log n +1$, $k=1,2,\cdots$, under the
condition that there exit two positive numbers $a>0$ and $d>0$ such
that $\Delta(G, a\log n)\leq d$. By Proposition 2.2, we know $m(G,
i, ka\log n)\leq ka\log n \Delta(G, a\log n)+ \delta_i\leq ka\log n
d+ \delta_i$, hence, $\delta(G, i, ka\log n)\leq d$ follows from the
definition $\delta(G,i,l)=(m(G,i,l)-\delta_i)/l$. By Proposition
2.3, it's sufficient to show $|S(T_{saw(i)}, i, ka\log n+1)|\leq
\delta_i(\delta(G, i, ka\log n)-1)^{ka\log n}\leq
\delta_i(d-1)^{ka\log n}=\delta_i(d-1)^{t-1}$. This is exactly what
we need. \ \ \ \ $\Box$\\

Next we will proceed to prove Theorem 2.2, we still use the
recursive formula but with another form. The technique used is a
well known method, Lipchitz approach. A `path' version of it will be
presented, which allow us to bound the `external field' with maximum
average degree. Before presenting it, we need some notation for
simplicity. Let $T=(V,E)$ be a tree rooted at $0$ with vertices
${0,1,2,\cdots,n}$, edge set $E$ and $TSSHC$ on it. For each edge
$(i,j)\in E$, recall the notation in main results,
$a_{i,j}=e^{\beta_{ij}(+,+)}$, $b_{i,j}=e^{\beta_{ij}(+,-)}$,
$c_{i,j}=e^{\beta_{ij}(-,+)}$, and $d_{i,j}=e^{\beta_{ij}(-,-)}$.
Let $M_{ij}=c_{ij}-d_{ij}$, $N_{ij}=a_{i,j}-b_{i,j}$ Define
\begin{displaymath}
f_{ij}(x)=\frac{M_{ij}x+d_{ij}}{N_{ij}x+b_{ij}},\ \ \ \
h_{ij}(x)=\frac{a_{ij}d_{ij}-b_{ij}c_{ij}}{(M_{ij}x+d_{ij})(N_{ij}x+b_{ij})}.
\end{displaymath}
Recall\\$\alpha_{\max}=\max\limits_{(i,j)\in
E}\{\beta_{ij}(-,-)-\beta_{ij}(+,-),\beta_{ij}(-,+)-\beta_{ij}(+,+)\}$,\\
$\alpha_{\min}=\min\limits_{(i,j)\in
E}\{\beta_{ij}(-,-)-\beta_{ij}(+,-),\beta_{ij}(-,+)-\beta_{ij}(+,+)\}$,
$\gamma_{ij}=\max\{\frac{|b_{ij}c_{ij}-a_{ij}d_{ij}|}{a_{ij}c_{ij}},
\frac{|b_{ij}c_{ij}-a_{ij}d_{ij}|}{b_{ij}d_{ij}}\}$,
$\gamma=\max_{(i,j)\in E}\{\gamma_{ij}\}$. For any $i\in V$, let
$T_i$ denote the subtree with $i$ as its root and $Z(i)$  be the
TSSHC induced on $T_i$ by $T$. Recall
$B_{i}=\frac{h_{i}(+)-h_i(-)}{2}$ is the `external field', denote
$\lambda_{i}=e^{-2B_i}$, and let $\Gamma_{ij}$ be the unique
self-avoiding path from $i$ to $j$ on $T$.\\

\textbf{Lemma 3.3} \emph{For any $(i,j)\in E$, $\max\limits_{x\in[0,1]}|h_{ij}(x)|\leq \gamma_{ij}$ }\\

\textbf{Proof:} The proof is technique and left to the appendix.\\

With the above notations, we present a `path' version Lipchitz
approach.\\

 \textbf{Lemma 3.4}\emph{ Let
$\Lambda\subset V$ , $\zeta_{\Lambda}$ and $\eta_{\Lambda}$ be any
two configurations on $\Lambda$. Let $\Theta=\{i: \zeta_i\neq
\eta_i, i\in\Lambda\}$, $t=d(0,\Theta)$ and $S(T,0,t)=\{i:
 d(0,i)=t, i\in T\}$. Then
\begin{displaymath}
\begin{split}
&|P_T(X_0=+|\zeta_{\Lambda})-P_T(X_0=+|\eta_{\Lambda})|\\
&\leq \gamma^{t}\sum\limits_{k\in S(T,0,t)}\prod\limits_{i\in
\Gamma_{0k} i\neq k}g_{i}(z_i)(1-g_{i}(z_i))
\end{split}
\end{displaymath}
where $g_{i}(x_i)=(1+\lambda_i\prod\limits_{(i,i_j)\in
T_i}f_{ii_j}(x_{ii_j}))^{-1}$ and $x_i$ is a vector with elements
$x_{ii_j}\in [0,1], i\in V$, and $z_i$ are constant
 vectors with elements in $[0,1]$. }\\

\textbf{Proof:} For any $i$ in $T$, let
$p^{\zeta_{\Lambda}}_{i}\equiv P_{T_i}(X_i=+|\zeta_{\Lambda_i})$ and
$R^{\zeta_{\Lambda}}_i\equiv
\frac{P_{T_i}(X_i=+|\zeta_{\Lambda_i})}{P_{T_i}(X_i=-|\zeta_{\Lambda_i})}$,
where $\zeta_{\Lambda_i}$ is configuration by restriction of
$\zeta_{\Lambda}$ on $T_i$. Then we have the following equality
\begin{displaymath}
\begin{split}
&p^{\zeta_{\Lambda}}_{0}=P_T(X_0=+|\zeta_{\Lambda})
=\frac{1}{1+\frac{P_T(X_0=-|\zeta_{\Lambda})}{P_T(X_0=+|\zeta_{\Lambda})}}
=\frac{1}{1+1/R^{\zeta_{\Lambda}}_0}\\
&=\frac{1}{1+\lambda_0\prod\limits_{(0,0_j)\in
T}\frac{c_{00_j}R^{\zeta_{\Lambda}}_{0_j}+d_{00_j}}{a_{00_j}
R^{\zeta_{\Lambda}}_{0_j}+b_{00_j}}}=\frac{1}{1+\lambda_0\prod\limits_{(0,0_j)\in
T}\frac{M_{00_j}p^{\zeta_{\Lambda}}_{0_j}+d_{00_j}}{N_{00_j}p^{\zeta_{\Lambda}}_{0_j}+b_{00_j}}}\\
&=g_0(x_0)
\end{split}
\end{displaymath}
where $x_0=(p^{\zeta_{\Lambda}}_{0_1},
p^{\zeta_{\Lambda}}_{0_2},\cdots,p^{\zeta_{\Lambda}}_{0_{\delta_0}})$.
First, note for any $x=(x_1,x_2,\cdots,x_q)$ and
$y=(y_1,y_2,\cdots,y_q)$, $q=\delta_0$, first order Taylor expansion
at $y$ gives that there exists a $\theta\in[0,1]$ such that
\begin{displaymath}
g_0(x)-g_0(y)=\nabla g_0(y+\theta (x-y))(x-y)^{T},
\end{displaymath}
where $(x-y)^{T}$ denotes the transportation of the vector $(x-y)$.
Calculate the $\frac{\partial g_0(x)}{\partial
x_i}$, we have \\
\begin{displaymath}
\begin{split}
\frac{\partial g_0(x)}{\partial
x_i}&=-\frac{\lambda_0\prod\limits^q_{j=1}f_{00_j}(x_{j})
(\frac{d\log(f_{00_i}(x_{i})}{dx_i})}
{(1+\lambda_0\prod\limits^q_{j=1}f_{00_j}(x_{j}))^{2}}\\
&=-g_0(x)(1-g_0(x))(\frac{M_{00_i}}{M_{00_i}x_i+d_{00_i}}-\frac{N_{00_i}}{N_{00_i}x_i+b_{00_i}})\\
&=g_0(x)(1-g_0(x))\frac{a_{00_i}d_{00_i}-b_{00_i}c_{00_i}}{(M_{00_i}x_i+d_{00_i})(N_{00_i}x_i+b_{00_i})}\\
&=g_0(x)(1-g_0(x))h_{00_i}(x_i).
\end{split}
\end{displaymath}
Hence, there's exits $\theta_0\in [0,1]$ such that
\begin{displaymath}
\begin{split}
|p^{\zeta_{\Lambda}}_{0}-p^{\eta_{\Lambda}}_{0}|&\leq
\sum\limits_{j=1}^{q}|g_0(z_0)(1-g_0(z_0))h_{00_j}(x_j)||p^{\zeta_{\Lambda}}_{0_j}-p^{\eta_{\Lambda}}_{0_j}|\\
&\leq
\sum\limits_{j=1}^{q}g_0(z_0)(1-g_0(z_0))\gamma_{00_j}|p^{\zeta_{\Lambda}}_{0_j}-p^{\eta_{\Lambda}}_{0_j}|\\
&\leq \gamma
\sum\limits_{j=1}^{q}g_0(z_0)(1-g_0(z_0))|p^{\zeta_{\Lambda}}_{0_j}-p^{\eta_{\Lambda}}_{0_j}|
\end{split}
\end{displaymath}
where
$z_0=p^{\eta_{\Lambda}}_{0}+\theta_0(p^{\zeta_{\Lambda}}_{0}-p^{\eta_{\Lambda}}_{0})$
and the second inequality follows by Lemma 3.3. Now repeat the
procedure for
$|p^{\zeta_{\Lambda}}_{0_j}-p^{\eta_{\Lambda}}_{0_j}|$,
$j=1,2,\cdots,q$, it is easy to see that the summation is over all
the self-avoiding paths emitting from the root $0$. For each path
$\Gamma$, if the end point of $\Gamma$ is a leave $j$ with
$d(0,j)\leq t-1$ or there is a vertex $i$ on $\Gamma$ with
$d(0,i)\leq t-1$ being fixed, the contribution of the path to the
summation is zero since
$p^{\zeta_{\Lambda}}_{i}-p^{\eta_{\Lambda}}_{i}=p^{\zeta_{\Lambda}}_{j}-p^{\eta_{\Lambda}}_{j}=0$.
Hence the remaining path with length $t$ is  in the set
$\{\Gamma_{0k}: k\in S(T,0,t) \}$. This completes the proof of lemma
3.4. \ \ \ \ $\Box$\\

In order to prove the Theorem 2.2, we need the following lemma.\\

\textbf{Lemma 3.5} \emph{ Let $\lambda_i\geq0$, $i=1,2,\cdots,n$.
Then
\begin{displaymath}
\prod\limits_{i=1}^n(1+\lambda_i)\geq
(1+\sqrt[n]{\prod\limits^n_{i=1}\lambda_i})^n.
\end{displaymath}}\\

\textbf{Proof:} The proof is technique and left to the appendix.\\

With Lemma 3.4 and 3.5, it is sufficient to prove Theorem 2.2.\\

\textbf{Proof of Theorem 2.2:} Following the notation of Lemma 3.4,
let $s=|S(T,0,t)|$,
we have \\
\begin{displaymath}
\begin{split}
|p^{\zeta_{\Lambda}}_{0}-p^{\eta_{\Lambda}}_{0}|&\leq
\gamma^{t}\sum\limits_{k\in S(T,0,t)}\prod\limits_{i\in \Gamma_{0k}
i\neq k}g_{i}(z_i)(1-g_{i}(z_i))\\
&\leq s\gamma^{t} \max\limits_{\substack{k\in
S(T,0,t)}}\prod\limits_{i\in \Gamma_{0k} i\neq
k}g_{i}(z_i)(1-g_{i}(z_i))\\
&\leq s\frac{\gamma^{t}}{4}\max\limits_{\substack{(0,0_j)\in T\\k\in
S(T,0,t)}}\prod\limits_{i\in \Gamma_{0_jk} i\neq
k}g_{i}(z_i)(1-g_{i}(z_i)).
\end{split}
\end{displaymath}
By Lemma 3.5, for each $\Gamma_{0_jk}$, $(0,0_j)\in T$, $k\in
S(T,0,t)$,
\begin{displaymath}
\begin{split}
&\prod\limits_{i\in \Gamma_{0_jk} i\neq
k}g_{i}(z_i)(1-g_{i}(z_i))\\&=\prod\limits_{i\in \Gamma_{0_jk} i\neq
k}\frac{\lambda_i\prod\limits_{(i,i_l)\in T_i}f_{ii_l}(z_{ii_l})}
{(1+\lambda_i\prod\limits_{(i,i_l)\in T_i}f_{ii_l}(z_{ii_l}))^{2}}\\
&\leq (\frac{r_{jk}} {(1+r_{jk})^2})^{t-1}
\end{split}
\end{displaymath}
where $r_{jk}=(\prod\limits_{i\in \Gamma_{0_jk} i\neq
k}\lambda_i\prod\limits_{(i,i_l)\in
T_i}f_{ii_l}(z_{ii_l}))^{1/(t-1)}$. A simple calculation gives that
$e^{\alpha_{\min}}\leq f_{ij}(x)\leq e^{\alpha_{\max}}$, for any
$(i,j)\in T$. Hence,
\begin{displaymath}
\begin{split}
e^{\alpha_{\min}(\delta(T,0,t-1)-1)}&\leq ( \prod\limits_{i\in
\Gamma_{0_jk} i\neq k}\prod\limits_{(i,i_l)\in
T_i}f_{ii_l}(z_{ii_l}))^{1/(t-1)}\\&\leq
e^{\alpha_{\max}(\delta(T,0,t-1)-1)}.
\end{split}
\end{displaymath}
Now we prove the exponential strong spatial mixing under assumption
of Theorem 2.2. Suppose $T$ is a self-avoiding tree of $G$.
$\delta(T,0,t-1)\leq \Delta(G,t-1)\leq d$ when $t=ka\log n+1$,
$k=1,2,\cdots$. If $B_{\min}> B(d,\alpha_{\max},\gamma)$, then
\begin{displaymath}
 \frac{\gamma(d-1)\exp(2B_{\min}-\alpha_{\max}(d-1))}
{(1+\exp(2B_{\min}-\alpha_{\max}(d-1)))^2}<1.
\end{displaymath}
Noting $s\leq \delta_0(d-1)^{t-1}$ and $(\prod\limits_{i\in
\Gamma_{0_jk} i\neq k}\lambda_i)^{1/(t-1)}\leq e^{-2B_{\min}}$ , now
we can see
\begin{equation}
\begin{split}
&|p^{\zeta_{\Lambda}}_{0}-p^{\eta_{\Lambda}}_{0}|\leq
s\frac{\gamma^t}{4}(\frac{r_{jk}} {(1+r_{jk})^2})^{t-1}\\
&\leq
\frac{\delta_{0}\gamma}{4}(\frac{\gamma(d-1)\exp(2B_{\min}-\alpha_{\max}(d-1))}
{(1+\exp(2B_{\min}-\alpha_{\max}(d-1)))^2})^{t-1}.
\end{split}
\end{equation}
The similar case holds for $B_{\max}< -B(d,-\alpha_{\min},\gamma)$.
This completes the proof.   \ \ \ \  $\Box$

Remark: As we point out in section II that if the graph $G$ is a
bounded degree graph with the maximum degree $d$, the condition in
Theorem 2.2 can be relaxed to $B_i>B(d,\alpha_{\max},\gamma)$ or
$B_i<-B(d,-\alpha_{\min},\gamma)$ for any $i\in V$. The reason for
this comes from the upper bound for $g_{i}(z_i)(1-g_{i}(z_i))$ in
the Lemma 3.4 since $\gamma(d-1)g_{i}(z_i)(1-g_{i}(z_i))<1$ for any
$i\in \Gamma_{0_jk}, i\neq k$ where ${(0,0_j)\in T,k\in S(T,0,t)}$.
We emphasize that one way to improve the condition by this method is
to carefully analyze the bound of $f_{ij}(x)$ for each iterative
step according to the range of $x$ since this will give better bound
for $g_i(x)$. We do not optimize the parameter here and do not know
whether dealing with the bound of $f_{ij}(x)$ carefully makes the
$B(d,\alpha_{\max},\gamma)$ or $-B(d,-\alpha_{\min},\gamma)$
optimally approximate the critical point of `external field' for
uniqueness of Gibbs measures either if there does exit one(note that
the critical points of `external field' for ferromagnetic and
antiferromagnetic Ising model are different on Cayley tree, an
infinite regular tree with the same degree for each vertex
\cite{Ge88}).

The proof of Theorem 2.3 will be shown in Section IV.
\section{Approximating Partition Function}
In the proof of Theorem 2.1 and 2.2, the calculation of the marginal
probability of the root yields a local recursive procedure.  If we
truncate the tree at height $t$, and then use the recursive method
to compute the marginal probability at root, it is easy to see the
complexity of this procedure is the number of vertices of truncated
tree. We now present the algorithm based on the above procedure and
self-avoiding
tree.\\
Let $G=(V,E)$ be a graph with vertices $V=\{1,2,\cdots,n\}$, edge
set $E$ and $TSSHC$ on it. Let  $\Phi_1$ denote the whole state
space ( which means $P_G(X_1=+|\Phi_1)=P_G(X_1=+)$), and
$\Phi_j=\{X_i=+, 1\leq i\leq
j-1 \}$, $2\leq j\leq n+1$.\\\\\\
\textbf{Algorithm for Partition Function $Z(G$)}\\\\
\textsf{\textbf{Input:} $G$ with the TSSHC, $\epsilon>0$ precision.\\
\textbf{Output:} $\widehat{Z(G)}$, the estimator of partition
function $Z(G)$.\\\\
For j=1:n\\
compute $\widehat{p_j}$, an estimator of conditional marginal
probability $p_j=P_G(X_j=+| \Phi_j)$, through self-avoiding tree
$T_{saw(j)}$ truncated at a certain height $t_j$ under the condition
$\Phi_j$ such that $(1-\frac{\epsilon}{2n})\leq
\frac{p_j}{\widehat{p_j}}\leq(1-\frac{\epsilon}{2n})$. (The initial
values of iteration at height $t_j$ are arbitrary nonnegative
numbers, if we adopt the recursive formula in the proof of Lemma 3.4
where $P_{T}(X_0=+|\zeta_{\Lambda})=g_0(x_0)$)\\\\
Output:$\widehat{Z(G)}=Z(G,\Phi_{n+1})\prod\limits^{n}_{i=1}\widehat{p_i}^{-1}$.}\\

With the above algorithm, it is enough to prove Theorem 2.3.\\

\textbf{Proof of Theorem 2.3:} First we show under the assumption of
the theorem, the Gibbs distribution exhibits exponential strong
spatial mixing. Since(by Proposition 2.4) $|V(T_{saw(i)},i,ka\log
n)|\leq |\max\limits_{j\in V}(V(T_{saw(j)},j,ka\log n))|^{k}\leq
(d-1)^{ka\log n}$ for any $i\in V$ and $k=1,2,\cdots$, we can obtain
the trivial bound of the number of vertices at height $ka\log n$,
that is ,  $|S(T_{saw(i)},i,ka\log n)\leq |V(T_{saw(i)},i,ka\log
n)|\leq (d-1)^{ka\log n}$. Let $t=ka\log n$ from the proof of
Theorem 2.1 and 2.2(see Formula (1) and (2)), substituting $(d-1)^t$
to $s$ in (1) (2), we get the exponential strong spatial mixing of
Theorem 2.3. Specifically, if $J<J_d$, the decay function
$f(t)=4J(d-1)((d-1)\tanh J)^{t-1}$ which corresponds to the
logarithmic form exponential strong spatial mixing, and if $J\geq
J_d$, $B_{\min}>B(d,\alpha_{\max},\gamma)$ or
$B_{\max}<-B(d,-\alpha_{\min},\gamma)$, the decay function has the
same form as in Theorem 2.2 except replacing $\delta_i$ by $d-1$. In
both cases, we suppose decay function $f(t)=be^{-ct}$ where $b$, $c$
are constant positive numbers independent of $n$, $t=ka\log n$,
$k=1,2,\cdots$. Through exponential decay property, it's sufficient
to show the above algorithm provides an FPTAS for $Z(G)$. Now we
check the output $\widehat{Z(G)}$ satisfying
$(1-\epsilon)\leq\frac{\widehat{Z(G)}}{Z(G)}\leq(1+\epsilon)$. Since
$p_j=\frac{Z(G,\Phi_{j+1})}{Z(G,\Phi_{j})}$, multiplying them gives
$Z(G)=Z(G,\Phi_{n+1})\prod\limits^{n}_{i=1}p_i^{-1}$. Hence,
$1-\epsilon\leq(1-\frac{\epsilon}{2n})^n\leq\prod\limits^{n}_{i=1}
\frac{\widehat{p_i}^{-1}}{p_i^{-1}}=\frac{\widehat{Z(G)}}
{Z(G)}\leq(1+\frac{\epsilon}{2n})^n\leq 1+\epsilon$. As we point out
previously that the complexity of the algorithm at each step is
$O(|V(T_{saw(j)},j,t_j)|)=O((d-1)^{t_j})$ when $t_j=ka\log n$,
$k=1,2,\cdots$, we only need to set $f(t_j)\leq
O(\frac{\epsilon}{2n})$ to promise $(1-\frac{\epsilon}{2n})\leq
\frac{p_j}{\widehat{p_j}}\leq(1-\frac{\epsilon}{2n})$ which requires
$t_j=O(\log n+\log(\epsilon^{-1}))$. Thus, the complexity of the
algorithm is $nO((d-1)^{O(\log
n+\log(\epsilon^{-1}))})=O(n^{O(1)}+n(\epsilon^{-1})^{O(1)})$, which
completes the proof. \ \ \ \ $\Box$

\section{Conclusion and Further Work}
We have shown that the Gibbs distribution of TSSHC on a `sparse on
average' graph $G=(V,E)$ with `maximum average degree' $d$ exhibits
the (exponential) strong spatial mixing when the absolute value of
`inverse temperature' $|J_{ij}|<J_d$ or the `external field' $B_i$
is uniformly larger than $B(d,\alpha_{\max},\gamma)$ or smaller than
$-B(d,-\alpha_{\min},\gamma)$, for any $(i,j)\in E$, $i\in V$. Here
$J_d$ is the critical point for uniqueness of Gibbs measure on a
infinite $d$ regular tree of Ising model, implying the condition for
inverse temperature is tight when restricting it on Ising model,
$B(d,\alpha,\gamma)$ is constant with parameter $d$, $\alpha$,
$\gamma$. It is not difficult to apply our results to
Erd$\ddot{o}$-R$\dot{e}$nyi random graph $G(n,d/n)$, where each edge
is chosen independently with probability $d/n$, since the average
degree in $G(n,d/n)$ is $d(1-o(1))$ while it contains many vertices
with degree $\log n/\log\log n$\cite{MS08}. As an application of
strong spatial mixing property, we present an FPTAS for partition
functions on a little modified sparse graphs, which includes
interesting bounded degree graph.

For future work, we expect to improve the condition on `external
field'. We have presented a way to improve it in the remark,
however, we believe the essential improvement needs other method.
Maybe the approach of analysis of the fixed point in\cite{Ke85}
works here.


\section*{Appendix}
\textbf{Proof of Lemma 3.3:} \\Since $M_{ij}x+d_{ij}\geq 0$ and
$N_{ij}x+b_{ij}\geq 0$, $\forall x\in[0,1]$, we need only to show
$\min\limits_{x\in[0,1]}w(x)=\min(a_{ij}c_{ij}, b_{ij}d_{ij})$,
where $w(x)=(M_{ij}x+d_{ij})(N_{ij}x+b_{ij})$. The case
$M_{ij}N_{ij}=0$ is trivial, so w.l.o.g. suppose
$M_{ij}N_{ij}\neq0$. Noting
$x_l=-\frac{d_{ij}N_{ij}+b_{ij}M_{ij}}{2M_{ij}N_{ij}}$ is an
extremum of $w(x)$ on $R$. There are three cases needed to be discussed.\\
Case 1. $M_{ij}N_{ij}<0$, then $w(x)$ reaches its minimum at
boundary. Then
$\min\limits_{x\in[0,1]}w(x)\leq\min(w(0),w(1))=\min(a_{ij}c_{ij},
b_{ij}d_{ij})$.\\
Case 2. $M_{ij}>0, N_{ij}>0$, then $x_l\leq 0$, $w(x)$ is increasing
on $[0,1]$, then $\min\limits_{x\in[0,1]}w(x)=w(0)=b_{ij}d_{ij}$.\\
Case 3. $M_{ij}<0, N_{ij}<0$, then $x_l\geq 1$, $w(x)$ is decreasing
on $[0,1]$, hence $\min\limits_{x\in[0,1]}w(x)=w(1)=a_{ij}c_{ij}$. \
\ \ \ $\Box$\\

\textbf{Proof of Lemma 3.5:}\\
\begin{displaymath}
\begin{split}
\prod\limits_{i=1}^n(1+\lambda_i)&=1+\sum\limits^n_{k=1}
(\sum\limits_{i_1<i_2<\cdots<i_k}\prod\limits^k_{j=1}\lambda_{i_j})\\
&\geq 1+\sum\limits^n_{k=1}(C^k_n(\prod\limits^n_{i=1}\lambda_i)^{\frac{C_{n-1}^{k-1}}{C^k_n}})\\
&=1+\sum\limits^n_{k=1}(C^k_n(\prod\limits^n_{i=1}\lambda_i)^{\frac{k}{n}})\\
&=(1+\sqrt[n]{\prod\limits^n_{i=1}\lambda_i})^n,
\end{split}
\end{displaymath}
where $C^k_n=\frac{n!}{k!(n-k)!}$. The first inequality uses the
arithmetic-geometric average inequality.
\end{document}